\documentclass[%
reprint,
 amsmath,amssymb,floatfix
]{revtex4-1}

\usepackage[english]{babel}

\makeatletter
\let\ORIbbl@fixname\bbl@fixname
\def\bbl@fixname#1{%
  \@ifundefined{languagealias@\expandafter\string#1}
    {\ORIbbl@fixname#1}
    {\edef\languagename{\@nameuse{languagealias@#1}}}%
}
\newcommand{\definelanguagealias}[2]{%
  \@namedef{languagealias@#1}{#2}%
}
\makeatother

\definelanguagealias{en}{english}
\definelanguagealias{En}{english}

\usepackage{graphicx}
\usepackage{dcolumn}
\usepackage{bm}
\usepackage{color}
\usepackage[normalem]{ulem} 

\def\>{\rangle}
\def\<{\langle}
\def\t{\dagger}

\begin{document}

\preprint{Advances in Physics X}

\title{Continuous measurements for control of superconducting quantum circuits}

\author{S. Hacohen-Gourgy}
 \affiliation{Department of Physics, Technion - Israel Institute of Technology, Haifa 32000 Israel}
 \email{shayhh@physics.technion.ac.il}
\author{L. S. Martin}%
\affiliation{Department of Physics, Harvard University, Cambridge, Massachusetts 02138, USA}

\date{\today}

\begin{abstract}
Developments over the last two decades have opened the path towards quantum technologies in many quantum systems, such as cold atoms, trapped ions, cavity-quantum electrodynamics (QED), and circuit-QED. However the fragility of quantum states to the effects of measurement and decoherence still poses one of the greatest challenges in quantum technology. An imperative capability in this path is quantum feedback, as it enhances the control possibilities and allows for prolonging coherence times through quantum error correction. While changing parameters from shot to shot of an experiment or procedure can be considered feedback, quantum mechanics also allows for the intriguing possibility of performing feedback operations during the measurement process itself. This broader approach to measurements leads to the concepts of weak measurement, quantum trajectories and numerous types of feedback with no classical analogues. These types of processes are the primary focus of this review. We introduce the concept of quantum feedback in the context of circuit QED, an experimental platform with significant potential in quantum feedback and technology. We then discuss several experiments and see how they elucidate the concepts of continuous measurements and feedback. We conclude with an overview of coherent feedback, with application to fault-tolerant error correction.

\end{abstract}

\maketitle

\section{\label{sec0}Introduction}

New technologies emerge once basic research acquires significant knowledge about a novel system, and yields sufficient control to enable new applications. Reaching such level of control is never simple in real systems, due to the multitude of degrees of freedom and the consequent noise. Information from the system can be used to feedback and correct the effects of noise. Feedback is used for stabilizing a system in a particular state or cause it to evolve in a desired manner, speeding up operations, or making them more efficient than the natural dynamics. This general approach applies to all systems, electrical, mechanical, and also to nascent quantum technologies.

A key property that gives quantum systems their technological potential is their coherence times, which are limited by noise from uncontrolled degrees of freedom. Quantum feedback is one of the only ways to actively reduce the noise and prolong coherence times in these systems. Thus, quantum feedback is a crucial component in quantum information processing tasks such as quantum meteorology and networking, and is a mandatory component in all fault-tolerant quantum computing proposals. However, quantum feedback control is more subtle than in classical mechanics, since feedback requires measuring the system, and measurement leads to wave function collapse.

In undergraduate quantum mechanics courses, we are taught that measurement leads to an instantaneous collapse of the wave function, a limit known as a projective measurement. This treatment almost always omits the study of open quantum systems, which creates the impression that decoherence, weak measurement and quantum trajectories are as mysterious as the measurement problem itself. In reality however, these phenomena follow directly from the basic axioms of quantum theory, and are completely independent of the philosophical issues regarding the interpretation of wave function collapse. The underlying connections are quite simple to state; weak measurement of a degree of freedom (position, spin, photon number \textit{etc.}) results when one interacts that degree of freedom with another system, and then applies a standard projective measurement to that secondary system. Decoherence is just weak measurement by the environment, or to be more precise, by a system that is inaccessible to the experimenter. A quantum trajectory of a state is the state of that system as it undergoes a continuous series of infinitesimally short weak measurements. All of these phenomena involve exchange of information, and in contrast to classical systems, the exchange of quantum information is necessarily accompanied by a disturbance in the quantum state, the so called measurement back-action. In this review, we will show how to treat this back-action as a control resource for quantum systems through the use of feedback. 

Feedback can be applied from shot to shot of an experiment and is useful that way for application such as quantum error correction. However, continuous measurements allow to fully harness the back-action as a control resource. This ability comes into play only if the experimenter has the ability to resolve the signals at time scales faster than the measurement induced decoherence. For example, in textbooks a photon detector is described as a click / no click output device, but a realistic device has intrinsic continuous dynamics. Having access to these dynamics opens up the control possibilities described in this review.

One of the more versatile systems that has recently emerged is the circuit-QED platform\cite{wallraff_strong_2004,paik_observation_2011,devoret_quantum_2014,langford_circuit_2013,schuster_schuster_2006,chow_quantum_2010}. On the one hand, circuit-QED is a platform for engineering artificial atoms and studying the physics of light-matter interactions. On the other hand, it is microwave circuit engineering, where the ‘atoms’ can be understood as non-linear oscillators. Engineering their impedance, coupling, and internal degrees of freedom forms a new quantum information processing technology. The experiments described here lie at the intersection of light-matter interaction, information processing, precision measurements, and system engineering. Specifically, we will discuss how to harness the back-action as a control resource within the context of continuous measurements in the circuit-QED platform. 

Section II introduces the circuit-QED platform. We detail a commonly used setup of a qubit in a cavity, and detail the workings of parametric amplification that affords the system with high precision continuous monitoring. We then discuss the formalism and analysis of such monitoring. In section III we introduce feedback. We discuss several experiments that demonstrate two types of control. The first type is feeding back on the stochastic back-action to stabilize or steer the system, commonly referred to as incoherent\cite{dong_quantum_2010} or measurement-based feedback and represented by (a) and (b) in Fig.\ref{fig:FBschemes}. The second is engineering the system such that the measurement back-action is evidently deterministic, driving the system to the desired state in an autonomous fashion, commonly referred as coherent feedback  and represented by (c) in Fig.\ref{fig:FBschemes}. It is now understood that the difference between these processes is physical and not fundamental, and which one can be applied is determined by the technological ability\cite{cruikshank_role_2017,balouchi_coherent_2017}. We conclude by expanding on the role of autonomous feedback in quantum error correction.

\begin{figure}
\begin{center}
{\includegraphics[width=1\linewidth]{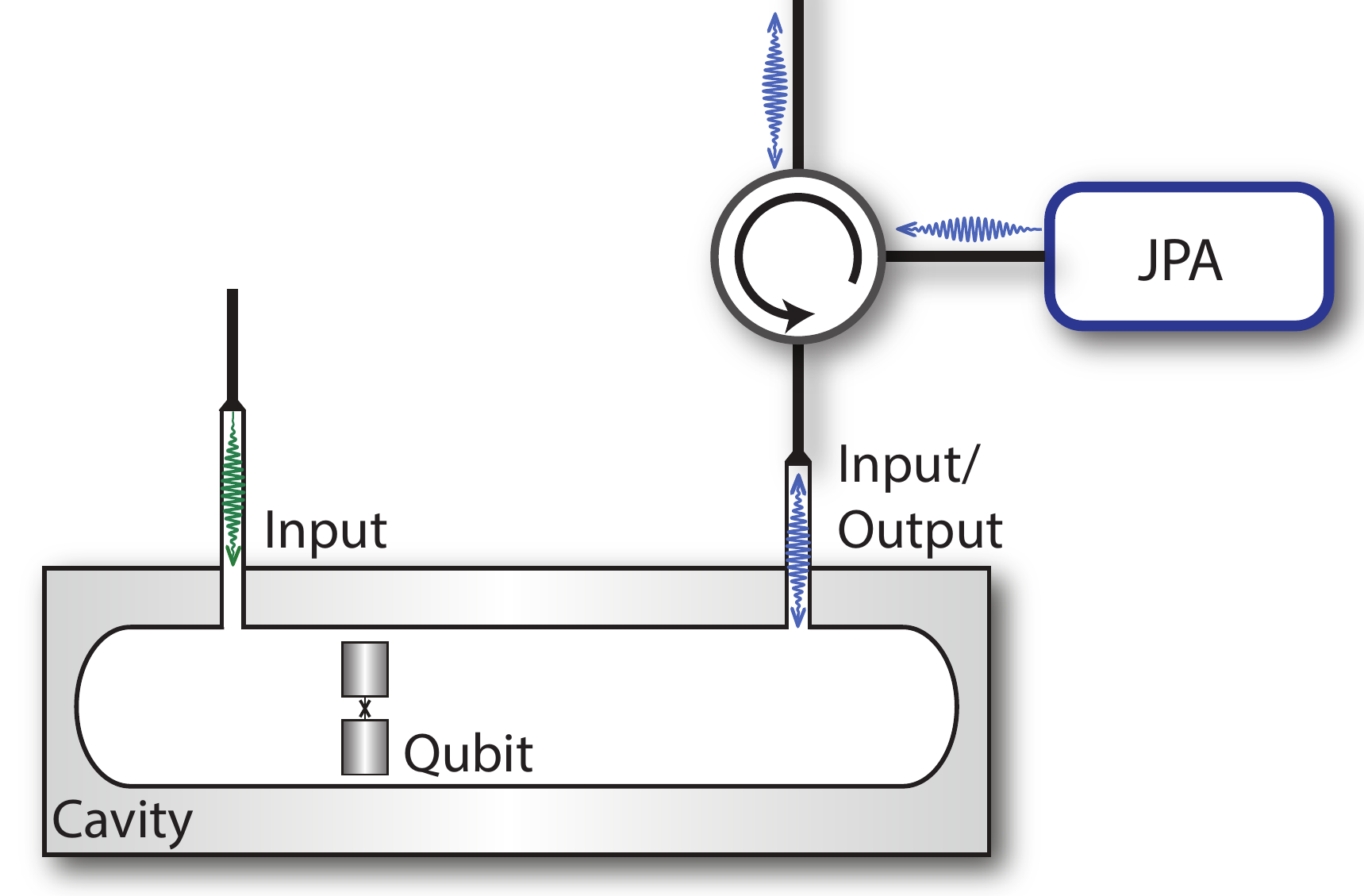}}
\caption{Illustration of a typical circuit-QED setup. A transmon qubit is placed inside a cavity. A Josephson parametric amplifier (JPA) monitors the cavity field through the strongly coupled port labeled `Input/Output'. In the dispersive limit, this setup continuously tracks the state of the qubit. An optional weakly coupled port labeled `Input' is also illustrated.}
\label{fig:Setup}
\end{center}
\end{figure}

\section{\label{sec1}Continuous measurements in circuit-QED}
The circuit-QED platform realizes a near-ideal quantum optical laboratory in the microwave regime. This platform evolved from early work showing macroscopic quantum tunneling in superconducting circuits\cite{widom_quantum_1979,devoret_measurements_1985,martinis_energy-level_1985, clarke_quantum_1988}, hence paving the way to engineering artificial atoms. In circuit-QED, electrical circuits are embedded with Josephson junctions that endow them with strong non-linearities. The circuits are coupled to superconducting resonators and controlled using microwave signals. The system can be understood as collections of engineered artificial atoms inside cavities. The effective light-matter interactions are orders of magnitude larger than in natural systems, such as cavity-QED platforms that use actual atoms \cite{haroche_exploring_2006}. This large interaction strength is the key to the system's versatility. 

In particular, it enables one to couple a tailored quantum system to a well-controlled reservoir like a transmission line, which itself can be monitored with high collection efficiency. This limit is essential for weak measurements, quantum trajectories and feedback.

Here we will focus on systems of up to a few artificial atoms coupled to an electromagnetic mode. The system, including the continuous monitoring, is very well described by the stochastic a master equation, where the Hamiltonian and the measurement rates are fully controlled. This realizes a unique test bed for quantum control using measurements.

\subsection{\label{sec11}Circuit-QED and the dispersive regime}
There are several excellent comprehensive reviews on circuit-QED~\cite{devoret_quantum_2014,langford_circuit_2013,schuster_schuster_2006,chow_quantum_2010}, reviewing the different types of qubits and their possibilities for coupling to resonators. We now describe the basic building blocks and the essential features required for the detection schemes presented in later sections.

The first step is to understand how to build artificial atoms using superconducting circuits. Superconductivity enables operation with practically no dissipation. At low temperatures, these circuits to have energy levels, just like other quantum systems. Notably, real atoms have energy levels with distinct spacings. This feature is important as it allows one to resonantly drive a specific transition without exciting any other transitions. The non-linear interactions, such as the Coulomb force, set the energy spectrum. Circuits consisting purely of ideal classical conductors, on the other hand, can realize only linear components, such as the capacitor and inductor. These linear components alone would limit us to constructing a linear, or harmonic, oscillator, which does not suffice for most quantum information applications. 

To realize nonlinearity with superconducting circuits, we can leverage the quantum dynamics of a circuit element known as a Josephson junction. The Josephson junction is composed of two superconductors that are weakly coupled, typically through a thin insulating barrier. The junction can carry a dissipationless current up to a certain limit called the critical current, which depends on the superconducting gap and the strength of the coupling. For the purposes of quantum circuit design, the Josephson junction can be modeled as a non-linear inductor with inductance $L(I)=\phi_0 /(2 \pi \sqrt{(I_c^2-I^2)})$, where $\phi_0$ is the flux quantum, and $I_c$ is the critical current of the junction. Details for designing and quantizing such circuits can be found in a review by U. Vool and M. H. Devoret~\cite{vool_introduction_2017}. While a variety of designs have been realized, we will focus on the transmon circuit~\cite{koch_charge-insensitive_2007}, which is currently the most widely used design due to its high coherence times and simple design consisting of a Josephson junction in parallel with a capacitor.
The transmon can be described as an anharmonic oscillator with negative anharmonicity of about 5\%. Typically the two lowest levels are utilized as the qubit, which can be described by the truncated Hamiltonian\cite{koch_charge-insensitive_2007},
\begin{equation}
H_q=\hbar \frac{\omega_q}{2} \sigma_z
\label{eq:qubitHam}
\end{equation}
where $\sigma_z$ is the Pauli spin operator and $\hbar \omega_q \simeq \sqrt{8 E_J E_C} $ is the energy difference between the two lowest levels, set by the circuit design. $2E_J$ is the maximum inductive energy that the Josephson junction can store, and is set by its material and geometric properties. $E_C$ is the charging energy associated with the displacement of one electron across the junction, which is set mainly by the capacitor shunting the Josephson junction.

An analogue of cavity-QED, in which an atom is coupled to an electromagnetic mode of a cavity, circuit-QED couples one or more superconducting qubits to a mode of a microwave cavity, \textit{i.e.} to a linear microwave resonator. The coupling of a natural atom with a dipole moment $\vec{p}$ to a single electromagnetic mode with field $\vec{E}$ at the position of the atom is,
\begin{equation}
H_{\mathrm{int}}=\vec{p} \cdot \vec{E}
\label{eq:dipole}
\end{equation}
where $\vec{E}=E_{\mathrm{ZPF}}(a^\dagger + a)\hat{n}$, $\hat{n}$ is the direction of the field, and $E_{\mathrm{ZPF}}$ is the amplitude of the zero-point fluctuations. Assuming a two-level atom, the interaction can be re-written as,
\begin{equation}
H_{\mathrm{int}}=\hbar g (a^\dagger + a) \sigma_x
\label{eq:TLSdipole}
\end{equation}
where $\hbar g \equiv \langle 1|\vec{p} \cdot \vec{n}|0 \rangle E_{\mathrm{ZPF}}$, and $\sigma_x$ is the Pauli spin operator, whose action is to flip the state of the two-level system. For the transmon circuit, which can be capacitively coupled to the field, the coupling would be of similar form, with $g$ depending also on the circuit design, and given by~\cite{koch_charge-insensitive_2007}
\begin{equation}
\hbar g = 2 \beta e \sqrt{\frac{\hbar \omega_{\mathrm{cav}}}{2 C_{\mathrm{cav}}}} \sqrt{\frac{1}{2}}\left(\frac{E_J}{8E_C}\right)^{1/4} 
\label{eq:TmonG}
\end{equation}
where $\beta$ is proportional to the capacitance between the transmon and the resonator, 
$C_\mathrm{cav}$ is the capacitance of the cavity resonator, and $\omega_{\mathrm{cav}}$ is the resonator transition frequency. Defining $\sigma^{\pm} \equiv 1/2 (\sigma_x \pm i \sigma_y)$, the interaction can be written as,
\begin{equation}
H_{\mathrm{int}}=\hbar g (a \sigma^+ + a^\dagger \sigma^- + a \sigma^- + a^\dagger \sigma^+)
\label{eq:BeforeJC}
\end{equation}
The last two terms exhibit rapidly-rotating time dependence in the interaction picture and can therefore be dropped in the typical case where $g \ll \{\omega_{cav},~ \omega_q\}$. This is called the rotating wave approximation (RWA), which yields the well-known Jaynes-Cummings Hamiltonian, a common starting point in describing circuit-QED and cavity-QED systems\cite{haroche_exploring_2006}. When resonant, the electromagnetic mode and qubit exchange energy at a rate $2g$, giving rise to vacuum Rabi oscillations.

An additional and extensively used approximation is the dispersive approximation. When the qubit is far detuned from the cavity mode by an amount $\Delta \equiv |\omega_q - \omega_{\mathrm{cav}}| \gg g $, direct energy exchange is highly suppressed. This limit is favorable for long qubit coherence times. To second order in $g/\Delta$ the interaction Hamiltonian $H_{\mathrm{int}}$ becomes,
\begin{equation}
H_{\mathrm{disp}}=\hbar \omega_{\mathrm{cav}} a^{\dagger} a + \hbar \frac{\omega_q}{2} \sigma_z + \hbar \chi a^\dagger a \sigma_z
\label{eq:DispHam}
\end{equation}
where $\omega_{\mathrm{cav}}$ and $\omega_q$ are re-normalized. $\chi=g^2/\Delta$ for a true two-level system, and for a transmon $\chi=g^2/\Delta (U/(U+\Delta))$, where $U$ is the anharmonicity. The difference in $\chi$ is due to the higher levels and the finite anharmonicity of the transmon~\cite{koch_charge-insensitive_2007}. Applying drives to the system is described by the following Hamiltonian,
\begin{align}
\label{eq:Hdrive}
H_{\mathrm{drive}} = \hbar(\alpha^{*}(t) a + \alpha(t) a^{\dagger}) 
\end{align}
Applying a drive at the qubit frequency allows for qubit control. 
Eqs. \ref{eq:DispHam} and \ref{eq:Hdrive} permit high fidelity qubit state readout. The last term of Eq. \ref{eq:DispHam} constitutes a qubit-state-dependent shift of the cavity resonance frequency by $\pm \chi$.
Applying a continuous drive at the bare cavity frequency via a transmission line coupled to the cavity generates a coherent state, whose phase thus depends on the state of the qubit.
Observing this phase shift using a high quantum efficiency amplifier is the key to the continuous monitoring of the state of the qubit.

\begin{figure}
\begin{center}
{\includegraphics[width=1\linewidth]{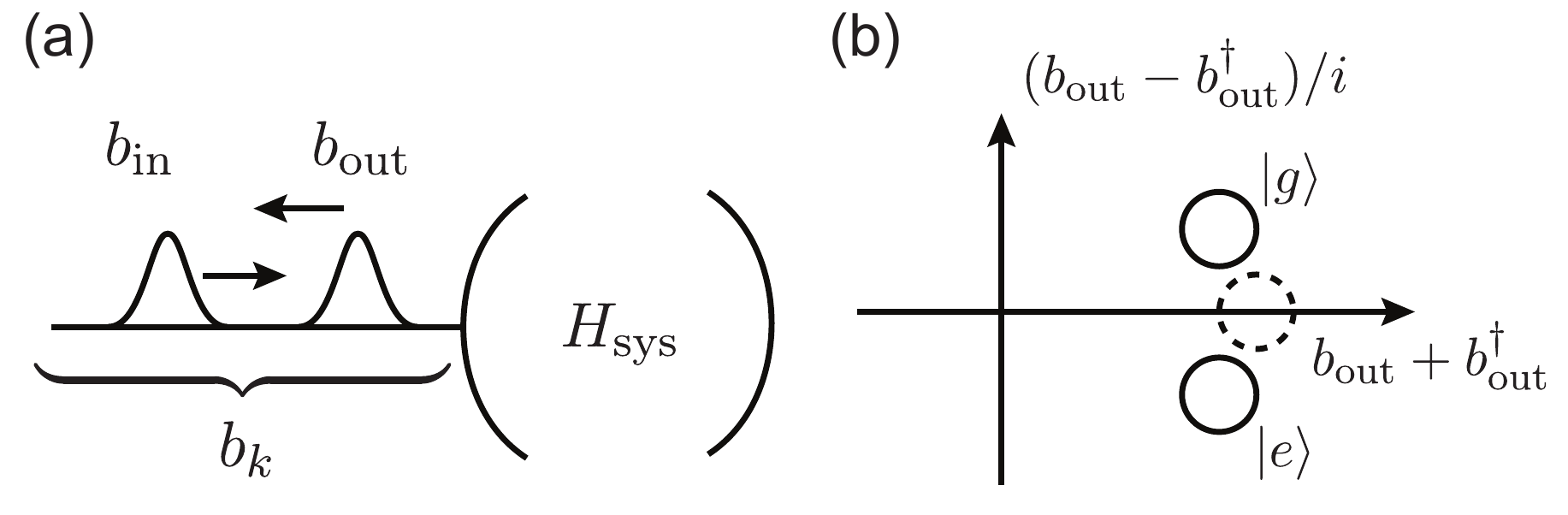}}
\caption{A transmission line coupled to a cavity. In input-output theory, we break the degrees of freedom on the transmission line into ingoing and outgoing modes. (b)  Coherent state representation of qubit readout. Dashed circle represents in the input state, while the solid circles represent the reflected output states conditioned on the qubit state.}
\label{fig:InputOutput}
\end{center}
\end{figure}

Let us first understand the measurement quantitatively, and then add amplification in the next subsection. We begin with the cavity input-output relations\cite{clerk_introduction_2010}, which replace Eq. \ref{eq:Hdrive} with a full quantum model of the transmission line\cite{clerk_introduction_2010}

\begin{equation} 
\dot{a} = \frac{i}{\hbar} [H_\text{sys}, a] - \frac{\kappa}{2} a(t) - \sqrt{\kappa} b_\text{in}[t]
\label{eq:IORelation}
\end{equation}

\begin{equation} \label{eq:IOBInOut}
b_\text{out}[t] = b_\text{in}[t] + \sqrt{\kappa} a(t)
\end{equation}

where $a$ is the cavity operator and $\kappa$ is the cavity damping rate. $b_{\mathrm{in(out)}}[t]$ are operators corresponding to ingoing(outgoing) temporal modes of the transmission line (Fig. \ref{fig:InputOutput}a). The system's amplitude \textit{i.e.} its quadrature operators $a \pm a^\t$ damp at a rate of $\kappa/2$ by emitting photons into the $b_\mathrm{out}$ modes. 

Dispersive qubit readout is an instructive application of the input-output relations. Suppose we have a qubit dispersively coupled to a cavity described by Eq. \ref{eq:DispHam}. One can use this Hamiltonian to perform a quantum non-demolition measurement of the qubit state, meaning that the measurement back-action commutes with the Hamiltonian, and in this case has no other action on the qubit but to project it into an eigenstate of the measurement operator $\sigma_z$. 

To see how this works, we can plug Eq. (\ref{eq:DispHam}) into Eqs. (\ref{eq:IORelation}) and (\ref{eq:IOBInOut}).
Note that putting $b_\text{in}$ in a coherent state has the same effect as adding a term proportional to $i(a^\t-a)$ to $H_\text{disp}$, which amounts to sending a sine wave at the cavity frequency in an experiment. We can compute the output field in the steady-state,
\begin{align} \label{eq:ReadoutCoherentStates}
\alpha_{g/e} &= \frac{\sqrt{\kappa}\epsilon}{\pm i \chi - \kappa/2} \\ \nonumber
\<b_\text{out}\> &= \epsilon \left( 1+ \frac{\kappa}{\pm i \chi - \kappa/2} \right) \\ \nonumber
&= \epsilon \frac{\chi^2 - \kappa^2/4 \mp i\chi \kappa}{\chi^2 + \kappa^2/4}.
\end{align}
where $\epsilon$ is the amplitude of the input drive, and $\alpha_{g/e}$ are the coherent states conditioned on the qubit being in ground of excited state. Thus the reflected phase of the output field depends on the qubit state, as illustrated in Fig. \ref{fig:InputOutput}b. Intuitively, this phase shift occurs because the $a^\dagger a \sigma_z$ term shifts the cavity resonance conditional on the qubit state.

If we can reach a signal-to-noise ratio much greater than 1, then we can also infer the specific measurement outcome in what is called a single-shot readout. For a large amount of noise we can increase $\epsilon$, or else the duration of the measurement, but both have their limitations. Readout duration is limited by the excited state lifetime of the qubit, $T_1$. Furthermore, the intercavity photon number $|\alpha|^2$ can only go so high before the dispersive approximation on which Eq. (\ref{eq:ReadoutCoherentStates}) is based breaks down. For a continuous measurement of the qubit, the low noise, or high quantum efficiency, is imperative to allow clear observation of the back-action of the measurement. To achieve low noise, we require a high quantum efficiency low temperature microwave amplifier.

\subsection{\label{sec11P}Parametric amplification in circuit-QED}

There are several circuit designs for high quantum efficiency parametric amplifiers in circuit-QED. While there are both travelling wave-based devices\cite{macklin_near_2015,obrien_resonant_2014,planat_photonic_2019}, and 

resonant structure-based devices\cite{bergeal_phase-preserving_2010,castellanos-beltran_amplification_2008,hatridge_dispersive_2011,frattini_optimizing_2018}, their operation and the underlying physics are relatively similar. Here we focus on a resonant circuit amplifier, the Josephson parametric amplifier (JPA). A JPA circuit consists of a Superconducting Quantum Interference Device (SQUID) shunted by a capacitor. A SQUID is a superconducting loop interrupted by two Josephson junction. The SQUID acts as a nonlinear inductance which increases with current, just like a single Josephson junction. The inductance can be increased either by applying magnetic flux through the SQUID loop, thus producing a circulating current in the SQUID, or by driving current across the SQUID directly. Thus the resonance frequency of the circuit can be modulated by either a flux-pump or a current-pump. A static flux-bias $\Phi_\text{dc}$ allows the JPA operating frequency to be tuned by up to an octave in some devices. A dynamic modulation of the JPA resonance frequency drives a parametric process in which a pump photon at $\omega_p=2\omega_0$ associated with the modulation is converted to a pair of photons, the signal ($\omega_s$) and idler ($\omega_i$) photons with $\omega_p=\omega_s+\omega_i$

A similar process also exists in which one drives an AC current through the JPA rather than the SQUID, called current pumping.
Comprehensive derivations of JPA amplification and squeezing and a derivation of the internal and output field of a JPA can be found in \cite{gardiner_quantum_2004,beltran_development_2011, clerk_introduction_2010}. 
 
For our purposes we can understand the JPA action through the squeezing operator, which we take as our starting point for brevity
\begin{align} \label{eq:SqueezingOpIdentity}
S(z) = \exp\left[ \frac{1}{2} \left(z^* a^2 - z a^{\t 2}\right) \right].
\end{align}
where $z=r e^{i \phi}$ is the squeezing parameter. The presence an $a^{\dagger 2}$ term represents creation of pairs of photons as described above. Eq. (\ref{eq:SqueezingOpIdentity}) lets us compute the amplitude quadrature operators after evolution under the action of the amplifier,

\begin{align} \label{eq:SqueezingQuadratures}
S^\t(z) I_{\arg(z)} S(z) &= e^{-|z|}I_{\arg(z)} \\ \nonumber
S^\t(z) Q_{\arg(z)} S(z) &= e^{|z|}Q_{\arg(z)} \\ \nonumber
I_\phi &\equiv a e^{-i\phi/2} + a^\t e^{i \phi/2} \\ \nonumber
Q_\phi &\equiv (a e^{-i\phi/2} - a^\t e^{i\phi/2})/i
\end{align}
using $\cosh(r) \pm \sinh(r) = e^{\pm r}$. $I_\phi$ and $Q_\phi$ are quadrature operators rotated in phase space by $\phi$, which are exponentially amplified and squeezed respectively. The squeezed quadrature $I_\phi$ will have fluctuations less than that of the vacuum state. The Heisenberg uncertainty principle is saved by the fact that the conjugate $Q_\phi$ has larger fluctuations by the same factor, conserving the product of the uncertainties. Importantly, Eq. (\ref{eq:SqueezingQuadratures}) shows that the squeezing Hamiltonian acts as a linear amplifier of the $Q_\phi$ quadrature. The enhanced fluctuations of the output field correspond to amplified vacuum fluctuations of the input mode.

Using the input-output relations (Eq. \ref{eq:IORelation} and Eq. \ref{eq:IOBInOut}) we can see how an externally imposed field, such as a qubit readout signal, is amplified. If the amplifier is fed by a transmission line with electromagnetic modes $b_\mathrm{in}[\omega]$, then the output field in the signal mode is given by\cite{martin_quantum_2020}
\begin{align}
\label{eq:ParampIORelation}
\tilde{b}_\text{out}(\Delta) &= -G_S \tilde{b}_\text{in}(\Delta) + \underbrace{i e^{i\arg(G_S)} \sqrt{|G_S|^2-1}}_{G_I} \tilde{b}_\text{in}^\t(-\Delta)  \\ \nonumber
G_S &= \frac{\kappa^2/4 + \Delta^2 + |\lambda|^2}{(\kappa/2-i \Delta)^2 - |\lambda|^2}
\end{align}
where $\kappa$ is the cavity linewidth and $\lambda=\epsilon \phi_{\mathrm{ZPF}}$ is proportional to the product of the pump amplitude and the zero point fluctuations. $\Delta$ is the detuning from resonance frequency. This frequency dependence is achieved by solving for the dynamics through a Fourier transform. $G_S$ and $G_I$ are called the signal and idler gains and $b_\mathrm{in}(-\Delta)$ is the input idler mode. For $\Delta=0$, the signal and idler modes coincide. By computing the output quadratures, one can easily verify that the output mode is simply a squeezed copy of the input. This scenario is referred to as phase-sensitive amplification, because whether or not a signal is amplified depends on its phase relative to the squeezing axis. 
For $\Delta \neq 0$, both input quadratures are amplified, but vacuum fluctuations from the idler mode $\tilde{b}_\text{in}(-\Delta)$ are mixed into the amplified signal, which contributes noise. In the large gain limit $\lambda\rightarrow \kappa/2$, we have $G_S \approx G_I$, indicating that ideally half of the signal consists of amplified idler vacuum fluctuations. This situation is called phase preserving amplification.

In summary, in phase sensitive mode one quadrature of the electromagnetic field is amplified, while the other is squeezed, whereas in phase preserving mode both quadratures are amplified, yet half a photon of noise must be added. When working in phase-preserving mode, this added noise is also part of the back-action, and needs to be taken into account for faithful tracking of the quantum system. When working in phase-sensitive mode, the axis of amplification will determine the back-action, and information is acquired only for the amplified quadrature of the field.

In practice, due to non idealities, a JPA monitoring a cavity provides typical quantum efficiencies of $\sim$50\%, with record values above 80\%\cite{eddins_high-efficiency_2019}.

\subsection{\label{sec12}Tracking the measurement back-action - Quantum trajectories}

\begin{figure}
\centering
{\includegraphics[width  =0.9\linewidth]{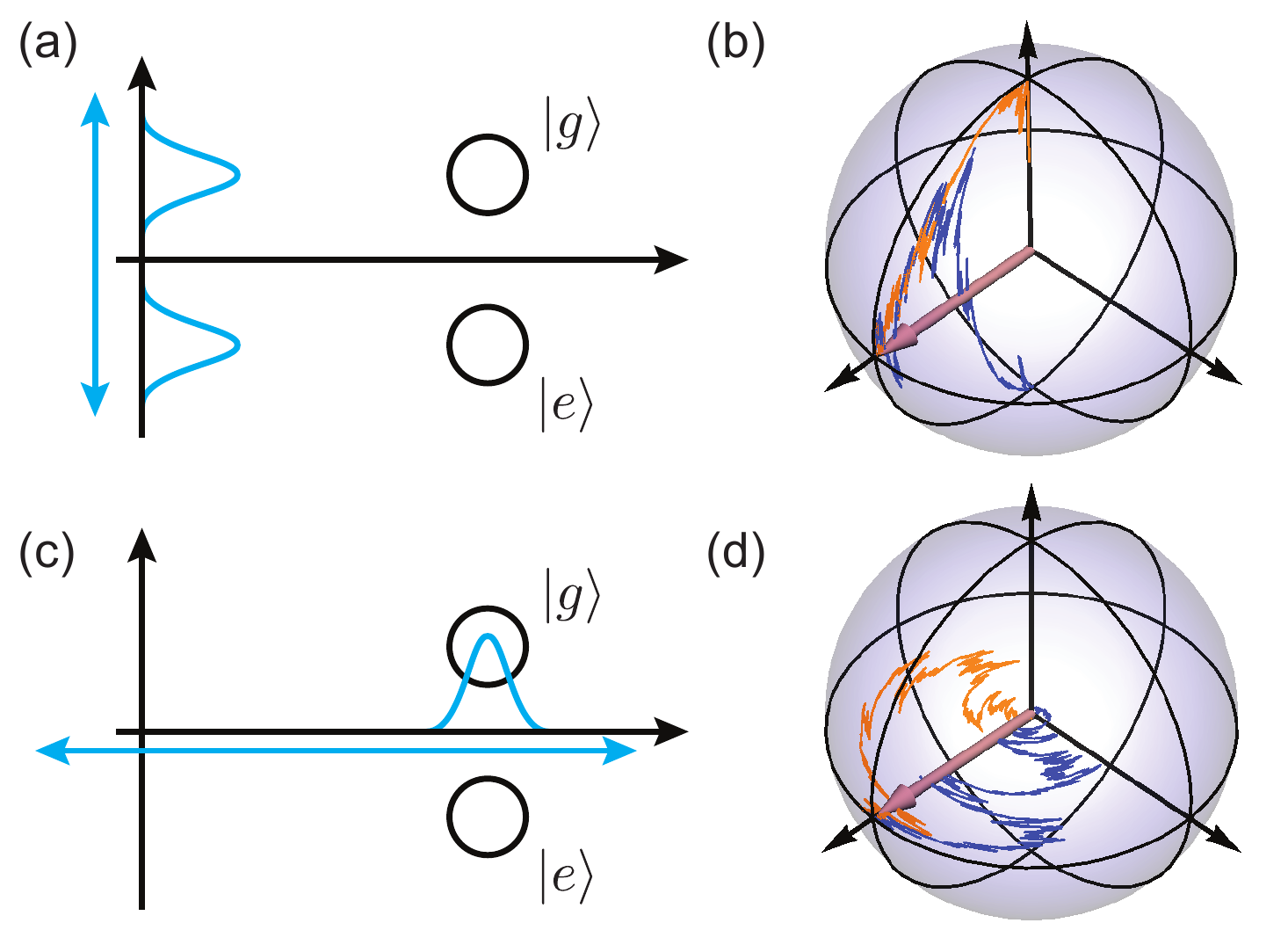}}
\caption{Simulated qubit trajectories via disperisive readout ($\eta=0.4$). (a) Measurement axis (blue) aligned with the quadrature containing qubit state information. Resulting trajectories (b) move toward the eigenstates of the measurement operator. (c) Measurement axis aligned perpendicular to the information-containing quadrature. Although qubit state information is hidden, measurement outcomes to the right end of the distribution project the cavity into a state with higher photon number. (d) Resulting quantum trajectories. Measurement-induced photon number fluctuations lead to stochastic precession around the measurement axis.}
\label{fig:QubitTrajectories}
\end{figure}

Tracking the back-action is a process wherein we continuously extract information about our system and update our best estimate of its state, \textit{i.e.,} calculate $\rho(t)$, the system's density matrix as function of time, conditioned on the measurement record. The continuous aspect can be understood if we divide time into discreet time steps $\Delta t$, and consider a weak interaction in each $\Delta t$ that is proportional to time $\Delta t$. A continuous measurement corresponds to taking $\Delta t$ to be infinitely small ($dt$). The back-action of the measurement at each time interval can be calculated by solving the stochastic master equation of the system\cite{jacobs_straightforward_2006}, which we describe at a high level below. Breaking up of the continuous process into discreet steps $\Delta t$ and applying a filter to infer information is similar to filtering in the context of classical control. This means that in many situations similar to the ones presented here, classical filtering results can be built upon and expanded to the quantum case\cite{doherty_feedback_1999}. The tracking process here is a special case of quantum filtering, where most of the results presented here can be understood as a Kalman filter\cite{zhang_kalman_2018}. While we focus primarily on circuit QED, the formalism of continuous measurement and feedback that we describe is applicable to many other quantum information platforms, particularly atomic cavity QED systems.

The simplest system that we will consider is a single transmon qubit coupled to a single cavity mode as described by Eq. (\ref{eq:DispHam}). The cavity is monitored by a JPA. The system is schematically illustrated in Fig. \ref{fig:Setup}. Typical design parameters for such a system are $\omega_{\mathrm{cav}}/2\pi$, $\omega_{\mathrm{q}}/2\pi$ in the 3 - 10 GHz range, and $\chi$ the dispersive shift and $\kappa$ the cavity linewidth are in the MHz range.

To model continuous measurement, first imagine a single temporal mode $b_\mathrm{in}$ in a transmission line, as depicted in Fig. \ref{fig:InputOutput}. After this mode reflects off of the cavity, it acquires a qubit-state-dependent phase shift, which in general entangles it with the qubit. If one then measures this mode, for instance by passing it through a JPA, then one acquires some information about the qubit state. Consequently, the qubit state is perturbed via the qubit-mode entanglement. This disturbance is the measurement back-action. However as we are only dealing with a single mode at a time, the amount of information gained is infinitesimal, as is the magnitude of the back-action. This fact suggests that a differential equation should exist to model the dynamics. Crucially, as the measurement outcomes are intrinsically random, the resulting differential equation should contain an element of randomness. Such a differential equation is called a stochastic differential equation.

Below we describe the stochastic master equation for a qubit that is continuously monitored by a JPA. Complete derivations specific to superconducting circuits may be found in \cite{gambetta_quantum_2008,martin_quantum_2020}. The stochastic master equation for a continuously monitored cavity is,
\begin{align}
\label{eq:CavSME}
d\rho = -i [H, \rho] dt 
+ \kappa \mathcal{D}[a] \rho ~dt + \sqrt{\kappa \eta} \mathcal{H}[a e^{i \phi}] \rho ~dW
\end{align}
where $\mathcal{D}[X]\rho = X\rho X^\dagger - (X^\dagger X\rho + \rho X^\dagger X)/2$, $\mathcal{H}[X]\rho = X\rho + \rho X^\dagger-\langle X\rho + \rho X^\dagger\rangle \rho$, $\kappa$ is the cavity decay rate, $\phi$ is the amplification axis of the JPA, $dW$ is a Gaussian distributed variable with variance $dt$, and $\eta$ is the JPA quantum efficiency. The first term in the equation we recognize from the Sch\"odinger equation formulated for density matrices. The second term is the dissipator, which damps the average photon number in the cavity at a rate $\kappa$. This term may also be familiar from the standard Lindblad master equation. The last term is the stochastic back-action part, which is the part responsible for updating $\rho$ according to the results of continuous monitoring. 

In this case of a monitored cavity, the phase $\phi$ sets the amplification axis of the field. If it is set along the qubit information axis (Fig. \ref{fig:QubitTrajectories}a) then qubit state information will be amplified. If the amplification axis is perpendicular (Fig. \ref{fig:QubitTrajectories}b), then qubit state information is squeezed.

Under the dispersive Hamiltonian Eq. \ref{eq:DispHam} and assuming that the qubit dynamics are slower than $\kappa$, tracking the cavity state gives rise to a continuous measurement of the qubit. We can see this using the polaron transformation, which maps Eq. \ref{eq:CavSME} to a stochastic master equation for the qubit\cite{gambetta_quantum_2008,hacohen-gourgy_quantum_2016},
\begin{align}
\label{eq:QubitSME}
d\rho &= -i [H, \rho] dt + \frac{\Gamma_D}{2} \mathcal{D}[\sigma_z] \rho ~dt \\ \nonumber & \sqrt{\eta \kappa |\alpha_e-\alpha_g|^2 \cos^2\phi} \mathcal{H}[\sigma_z] \rho ~dW \\ \nonumber & -i\frac{\sqrt{\eta \kappa |\alpha_e-\alpha_g|^2 \sin^2\phi}}{2} [\sigma_z,\rho] ~dW
\end{align}
where $\Gamma_D=2 \chi \mathrm{Im}(\alpha_g \alpha^*_e)$ is the measurement induced dephasing rate. The second term is information about the measurement operator $\sigma_z$, and thus the back-action pushes the state towards the poles of the qubit Bloch sphere. The third term is information about the qubit phase, and thus induces rotations around the $\sigma_z$ axis. Experimentally, $dW$ is given by the measurement record,
\begin{align}
\label{eq:dWmeas}
Vdt=\langle \sigma_z \rangle dt + \frac{dW}{\sqrt{2 \eta \Gamma_D}}.
\end{align}

In practice it is helpful to numerically solve the stochastic master equation using the corresponding Positive Operator Value Measure (POVM). The POVM can be understood as the generalization of a projection operator to weak measurements\cite{jacobs_straightforward_2006}. Taking $\eta=1$ for simplicity, and since $\sigma_z$ is Hermitian, Eq. (\ref{eq:QubitSME}) is equivalent to the following state update equation
\begin{align}
\label{eq:POVM}
\Omega(V) &= \exp \left[-\frac{\Gamma_D}{2} \left(V(t)-\sigma_{z}\right)^2 dt \right] \\ \nonumber
\rho(t+dt) &= \frac{\Omega \rho(t) \Omega^\dagger}{\text{Tr}[\Omega \rho(t) \Omega^\dagger]}. \\ \nonumber
\end{align}
The dynamics $\rho(t)$ for each iteration are referred to as a quantum trajectory, as coined by Carmichael\cite{carmichael_quantum_1993} almost three decades ago. There has been a plethora of research utilizing this tool since the first demonstration of quantum trajectories in circuit-QED by Murch \textit{et al.}\cite{murch_observing_2013}.
Fig. \ref{fig:QubitTrajectories} shows examples of quantum trajectories, simulated using Eq. \ref{eq:dWmeas} and \ref{eq:POVM}. Fig. \ref{fig:QubitTrajectories}b shows trajectories where the JPA axis is aligned to amplify qubit information, whereas Fig. \ref{fig:QubitTrajectories}d shows trajectories where the JPA axis is perpendicular to qubit information quadrature. In the latter case, the back-action does not push the state towards one of the $\sigma_z$ eigenstates. The dynamics are stochastic fluctuation around the $\sigma_z$ axis. This occurs since the JPA is squeezing all information about $\sigma_z$, yet amplifying information about cavity photon number fluctuations. The cavity photon number is coupled to the $\sigma_z$ operator through the dispersive term $\chi a^\dagger a \sigma_z$ (Eq. \ref{eq:DispHam}). The measured photon number fluctuations cause the dephasing, and hence the stochastic fluctuation around the $\sigma_z$ axis.

An important aspect for these measurements is the filter used to extract the trajectories from the data. Here we introduced the stochastic master equation and gave one solution for update of a monitored qubit, there are other methods of inferring the trajectories\cite{korotkov_quantum_2016,rouchon_efficient_2015,laverick_general_2020}. For example, Murch \textit{et al.}\cite{murch_observing_2013} used Bayes rule to update the observer's state of knowledge, and hence the density matrix. The probability for the qubit to be in state $i$ is given by $P(i|V_m)=P(i) P(V_m|i) / P(V_m)$, where $V_m=\int_{0}^{t} V(t')dt'$ is the integrated signal from the detector. Recognizing that the diagonal elements of $\Omega(V)^\dagger \Omega(V)$ are proportional to $P(V|i)$ (where $i$ labels the eigenvectors of $\sigma_z$), we can recognize that the Bayesian approach is closely related to the POVM-based state update.

Eq. \ref{eq:POVM} is valid for finite time intervals as well, provided that the measurement operator commutes with the Hamiltonian. In the general case, the Hamiltonian has to be known and taken into account, and update of the density matrix has to occur at time steps far smaller than the dephasing rate and in a time ordered fashion\cite{hacohen-gourgy_quantum_2016,weber_mapping_2014}. 
While most state update models are mathematically equivalent in the $dt \rightarrow 0$ limit, approximate heuristics can offer important advantages in terms of efficiency and robustness. For example, neural networks can learn and efficiently apply the correct state update in the presence of experimental imperfections and systematic uncertainties, even when the measurement and Hamiltonian dynamics do not commute\cite{flurin_using_2020}.

\begin{figure*}
\begin{center}
{\includegraphics[width=1\linewidth]{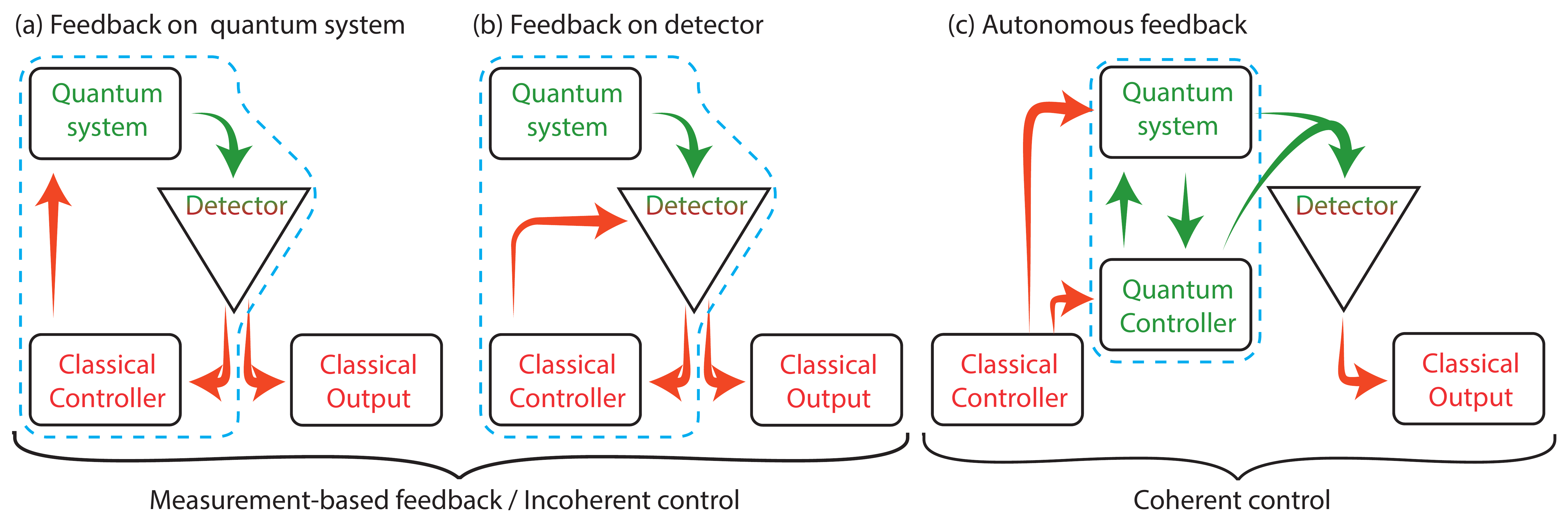}}
\caption{Schematics of quantum control. (a) Typical feedback scheme, where information is used to feedback directly to the quantum system. This situation is analogous to most feedback schemes in classical systems. (b) The controller feeds back on  the detector, and the quantum system is controlled through the back-action of the detector itself. (c) No classical controller. The feedback occurs at the quantum level. A larger Hilbert space is utilized, and although it is not always very apparent, it can be understood as a quantum controller feeding back on the target quantum system. This process (b) and (c) necessarily utilize the back-action as a control resource. Experiments are not restricted, and it is possible to utilize more than one type of feedback in a single setup.}
\label{fig:FBschemes}
\end{center}
\end{figure*}

\section{\label{sec2}Back-action and quantum control}
Going a step beyond simply tracking the evolution, we can use the measurement to actually influence the system through back-action. The change of the state of a quantum system as a consequence of the extraction of information can be used to steer the system in a desired way using a controller. In this respect, the back-action should be viewed as an additional control resource beyond Hamiltonian control, rather than a random disturbance. When a quantum system has access to projective measurements only, then the feedback discussed is necessarily a change of a control parameter from shot to shot of the experiment. With continuous measurements, we are interested in feedback during a single shot, meaning feeding back \textit{during} the readout itself. The control can occur in two ways. One, letting the back-action act in a stochastic way and feedback according to the acquired information, as illustrated in Fig. \ref{fig:FBschemes}a and Fig. \ref{fig:FBschemes}b, and two, designing the system such that the back-action operates in a deterministic fashion on the quantum level, as illustrated in Fig. \ref{fig:FBschemes}c. In an experiment, one does not have to choose a scheme, and these methods can be combined.

\begin{figure}
\centering
\includegraphics[width = 1\linewidth]{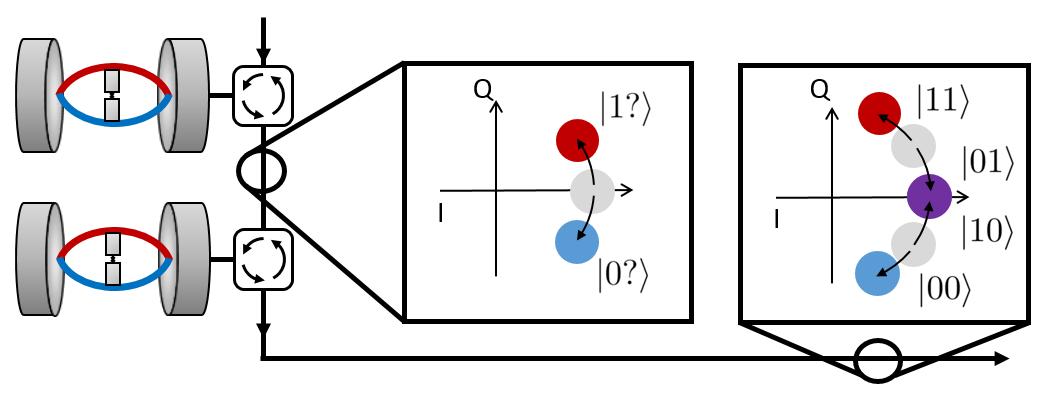}
\caption{A graphical illustration of the bounce-bounce remote entanglement experiment performed in \cite{roch_observation_2014}. A coherent measurement tone interacts sequentially with two cavities, each housing a superconducting transmon qubit. The final coherent state picks up two conditional phase shifts, and cannot distinguish between the $|01\>$ and $|10\>$ states.}
\label{fig:BounceBounce}
\end{figure}

Before diving into feedback, let us first look at a case where the back-action can be useful even without feedback of any sort, albeit only in a probabilistic fashion. Such an experiment, performed in circuit-QED is illustrated in Fig. \ref{fig:BounceBounce}\cite{roch_observation_2014}. The authors placed two qubits in two identical cavities and monitored them using a dispersive measurement. The setup is such that the signal reflects off of one cavity, bounces to the next one and then reflects to a JPA. The coherent tone picks up a phase shift conditioned on the qubit states. If the phase shifts are equal in magnitude, then the final output signal cannot distinguish the $|01\>$ and $|10\>$ states. Performing a homodyne measurement of this output field implements measurement of the operator\cite{motzoi_continuous_2015}
\begin{align}
M = \sqrt{\frac{\Gamma}{2}}\frac{\sigma_{z,1} + \sigma_{z,2}}{2}
\end{align}
where $\Gamma$ denotes the effective measurement rate. The above operator is called a half-parity measurement, and it is degenerate in the $|01\>$, $|10\>$ subspace. This can be used to probabilistically generate entanglement by first preparing the separable uniform superposition state $|\psi_0\> = \frac{1}{2}(|00\>+|01\>+|10\>+|11\>)$ and then projectively measuring $M$. Half of the time, the measurement outcomes $\pm1$ occur (dropping unitful quantities) and we collapse into the separable states $|00\>$ or $|11\>$. The other half of the time, the system collapses into the $|01\>$, $|10\>$ subspace while retaining its coherence within that subspace. The result is a Bell state $|\psi^+\> \equiv \frac{1}{\sqrt{2}} (|01\>+|10\>)$. The above process, and many like it, are intrinsically probabilistic. Even if implemented ideally, the success rate cannot exceed 50\%, which means that the unheralded state $\bar{\rho} = \frac{1}{4}|00\>\<00| + \frac{1}{4}|11\>\<11| + \frac{1}{2}|\psi^+\>\<\psi^+|$ has no entanglement. In the following section we will see how feedback can deterministically entangle the qubits in this example, and in general allow us to fully utilize the back-action as a  control resource.

\subsection{\label{sec21}Stochastic back-action and measurement-based feedback}
Measurement-based feedback is a broad topic, and the term carries many potential meanings, from state initialization to quantum error correction. The first systematic treatment of feedback in the context of continuous measurement was carried out by Wiseman and Milburn\cite{wiseman_quantum_1994, wiseman_quantum_1993}. A key contribution of these works was the introduction of a feedback master equation in the context of diffusive measurements such as Eq. (\ref{eq:CavSME}). The feedback master equation models continuous measurement supplemented with a control Hamiltonian, analogous to proportional feedback in classical control theory. In this section, we introduce this master equation and then give several example applications.\cite{martin_single-shot_2019,martin_quantum_2020}

For generality, we allow for multiple measurement operators and feedback Hamiltonians. The latter choice allows one to model feedback in which the form of the control Hamiltonian can depend on the measurement record, not just its overall magnitude. Consider a continuous measurement of a set of operators $M_i$ with efficiencies $\eta_i$, and each having a measurement record $V_i=\sqrt{\eta_i}\langle M_i+M_i^\dagger \rangle dt + dW_i$. Now consider continuously applying a set of control Hamiltonians $H_j$ each with strength $B_j dt + \sum_i A_{ij} dW_i$. The feedback master equation, obtained by averaging the measurement record, is given by\cite{wiseman_quantum_2010,martin_single-shot_2019},
\begin{align} \label{eq:FeedbackMasterEq}
\frac{d\rho}{dt}&=-i\sum_j B_j [H_j,\rho] \\ \nonumber &+ \sum_i[\mathcal{D}[M_i]\rho-i\sqrt{\eta}[\widetilde{H}_i,\mathcal{H}[M_i]\rho]+\mathcal{D}[\widetilde{H}_i]\rho] \\ \nonumber \widetilde{H}_i &\equiv \sum_j A_{ij} H_j
\end{align}
Locally optimal feedback amplitudes $A_{ij}$ and $B_j$ can be determined by maximizing the expectation value with respect to some target operator, and can be found in Ref\cite{martin_single-shot_2019}. 

Eq. (\ref{eq:FeedbackMasterEq}) is also referred to as the Wiseman-Milburn equation\cite{wiseman_quantum_2014}. It serves as the equations of motion for a system undergoing continuous measurement, proportional feedback and evolution under a time-dependent Hamiltonian (via the $B_j[H,\rho]$ term, though this term also contains a contribution from proportional feedback). The $\mathcal{D}[\widetilde{H}_i]$ term arises from random, noisy application of $H_j$ under feedback, which explains why it is present even if the measurement efficiency goes to zero. The term proportional to $\sqrt{\eta}$ derives from correlation between the measurement back-action and the feedback operations applied as a result, so it is most readily interpreted as the feedback term. We now consider several circuit-QED experiments that can be understood within this context.

\textit{Feeding back on the qubit} - We now consider the example of a single, resonantly driven transmon qubit. The qubit population oscillates between the ground and excited states at the Rabi frequency, which is proportional to the amplitude of the resonant drive. A typical experiment for observing Rabi oscillations would projectively measure the state of the qubit many times with varying durations of the resonant driving. The averaged oscillations will decay with a time constant that reflects the environmentally induced dephasing. 

If we now ignore the environment, yet add a continuous measurement during Rabi driving, the same decaying oscillation will appear. The decay rate will now correspond to the measurement rate. We can understand the measurement as an effective environment. However unlike environmental dephasing, continuous measurement enables tracking of the back-action. As suggested by Korotkov\cite{korotkov_persistent_2011} and demonstrated experimentally by Vijay et al.\cite{vijay_stabilizing_2012}, we can stabilize Rabi oscillations by adjusting the amplitude of the drive in proportion to the back-action. In simple illustrative terms, if the back-action kick due to the measurement was such that the qubit rotated faster (slower) than it should have, then lowering (increasing) the drive amplitude momentarily will put the qubit back on track. The information on the back-action kick is contained in the measurement signal. The beautiful simplicity of this feedback protocol allowed it to be implemented using analog mixing of the returning signal with the applied Rabi drive. The authors showed that the ensemble Rabi oscillations can be stabilized indefinitely. They also showed that the amplitude of the oscillations directly corresponds to the quantum efficiency of the JPA. This emphasizes the notion that the feedback is as good as its ability to acquire information about the measured system.

We now return to the example of the half parity measurement of two qubits. By adding feedback during the continuous measurement process, a Bell state can be produced deterministically.  
To do so, feedback should correct for measurement perturbations toward the undesired measurement eigenstates states $|00\>$ and $|11\>$. We can rotate between these states and $|\psi_0\>$ with a $\pi/2$ rotation of $H = (\sigma_{y,1}+\sigma_{y,2})/2$, and from $|\psi_0\>$ we can probabilistically prepare the target state via measurement. Intuitively, infinitesimal rotations of $H$ can steer the measurement outcome toward the desired subspace continuously. Initializing the system in $|\psi_0\>$ and assuming $\eta=1$ for simplicity, the state evolves under proportional feedback as \cite{martin_deterministic_2015}
\begin{align} \label{eq:HPFSolution}
    |\psi(t)\> = \frac{1}{\sqrt{2}}\left(e^{-\Gamma t/4}|\phi^+\> + \sqrt{2-e^{-\Gamma t/2}}|\psi^+\>\right)
\end{align}
where $|\phi^+\> \equiv \frac{1}{\sqrt{2}} (|00\>+|11\>)$. We have $|\psi(0)\> = |\psi_0\>$ and $|\psi(t\gg 1/\Gamma)\> = |\psi^+\>$, indicating that we can deterministically prepare an entangled state from a separable state using feedback. A detailed derivation including a proof of global optimality of this resulting protocol is given in Ref. \cite{martin_what_2017}. Note that $|\psi(t)\>$ only depends on time, and not on any function of the measurement record. This occurs because the stochastic terms of the master equation, one coming from measurement back action and one from feedback, cancel exactly. This cancellation is somewhat surprising, given that feedback is local while measurement is not.

\textit{Feeding back on the measurement operator} - We now consider the situation in which we vary the measurement operator conditional on the measurement outcome, as illuastrated in Fig. \ref{fig:FBschemes}b. In the broadest sense, we can think of adaptive measurement as a kind of `POVM-generating machine,' which widens the range of observables that may be accessed by a detector. 

As an instructive example, we look at the experiment demonstrating the adaptive phase measurement of a single microwave photon\cite{martin_implementation_2020}. Although no quantum mechanical operator corresponds to electromagnetic phase\cite{busch_are_2001} in the sense of a standard projective measurement, the canonical phase measurement defines the quantum-mechanically ideal POVM for measurement of phase,
\begin{align} \label{eq:CanPhase}
\Omega_\phi &= |\phi\>\<\phi| \\ \nonumber
|\phi\> &\equiv \frac{1}{\sqrt{2\pi}} \sum_{n=0}^\infty e^{i\phi n} |n\>.
\end{align}
In this experiment, the system adaptively changes the measurement basis of a JPA during the arrival of a single microwave photon superposed with the vacuum. The purpose of feedback control is not to prepare a target state, as in the previous example, but rather to optimize what information the detector acquires. 

The experimental setup is the same as in Fig. \ref{fig:Setup}. Sideband cooling on the cavity\cite{murch_cavity-assisted_2012} was used to emit a single photon state. This is a form of autonomous feedback and will be discussed in the next section.
By modulating the sideband amplitude during photon emission, a photon with a flat modeshape was generated, which helped ameliorate the detrimental effects of feedback delay at the receiver. The phase $\Theta_\mathrm{true}$ was encoded by preparing the qubit in a superposition state of the form $(|0\rangle+e^{i\Theta_\mathrm{true}}|1\rangle)/\sqrt{2}$, which decayed by emitting the photonic state $(|0\rangle + e^{i\Theta_\mathrm{true}}|1\rangle)/\sqrt{2}$. To adapt the measurement basis, the JPA was pumped by a field-programmable gate array (FPGA), which served as a classical feedback controller. The total quantum efficiency was $\eta = 0.4$. The JPA measured the field amplitude via the quantum mechanical quadrature operator $a e^{-i\phi(t)} + a^\dagger e^{i\phi(t)}$, where $\phi(t)$ is the instantaneous phase of the parametric pump.

\begin{figure}
\centering
{\includegraphics[width=1\linewidth]{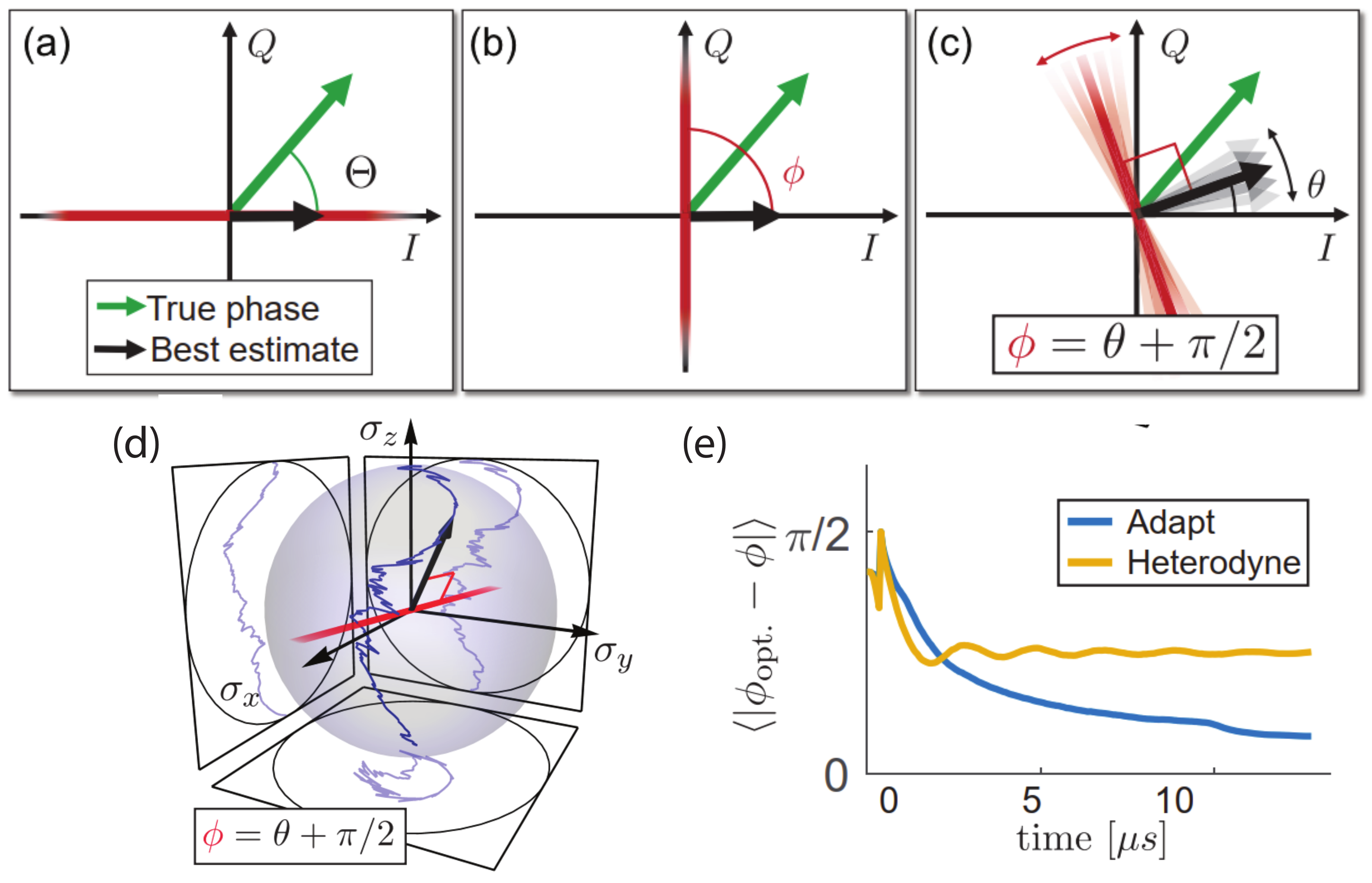}}
\caption{Measurement back-action and quantum trajectories. (a-c) Estimating and tracking state by changing measurement basis. Receiver attempts to maintain the phase measurement condition $\phi=\theta+\pi/2$. (d) A single adaptive-dyne quantum trajectory. The red right-angle bracket emphasizes orthogonality between the measurement axis and the state. (e) Quality of tracking for heterodyne and adaptivedyne. Adaptivedyne significantly outperforms the heterodyne and comes close to the ideal phase by $T=13 \mu s$.
}
\label{fig:back-action}
\end{figure}

To perform the canonical phase measurement on the incident field, the feedback controller continuously adapts the measurement axis $\phi(t)$ as the photon arrives at the receiver\cite{wiseman_adaptive_1995}. The measurement axis is chosen to maximize the acquisition of phase information as follows. Before the photon reaches the JPA, the receiver has no information and therefore chooses $\phi$ arbitrarily. Upon arrival of a portion of the photon, the JPA detects a small positive (or negative) fluctuation, which then informs the system that the true phase is likely oriented along (or opposite) the measurement axis (Fig. \ref{fig:back-action}a). At this point, any further measurement in this basis interrogates the amplitude of the incident field, which yields undesired photon number information. Ideally, the system would then rotate the measurement axis by 90 degrees (Fig. \ref{fig:back-action}b), so that a small deviation between the current best estimate of the phase $\theta(t)$ and the true phase $\Theta_\mathrm{true}$ would be detectable as a positive or negative fluctuation in the signal. As the photon continues to arrive, the feedback controller gains more information and updates the phase $\phi(t)$ to maximize sensitivity to phase (Fig. \ref{fig:back-action}c). If the phase measurement condition $\phi(t) = \theta(t) + \pi/2$ is maintained at all times, then the system acquires only phase information and implements a canonical phase measurement Eq. (\ref{eq:CanPhase}).

A unique feature of quantum feedback applied to a detector is that the presence of back-action offers a method to independently validate the receiver. This is done through the trajectories of the transmitter qubit itself. When the measurement axis is aligned with the best estimate of the phase ($\phi = \theta$), the resulting acquisition of amplitude information manifests as a random disturbance of the qubit state along the axis of decay. Conversely, when the phase measurement condition is satisfied ($\phi = \theta+\pi/2$), then only the phase of the qubit state is subject to noise. The performance of the receiver can be characterized through the dynamics of the qubit. This capability arises from entanglement between the qubit and its emitted photon. Fig. \ref{fig:back-action}d shows a single quantum trajectory under adaptivedyne detection, in which $\phi(t)$ is continuously adapted by the feedback controller. Fig. \ref{fig:back-action}e shows the difference between the ideal quadrature phase and the measured phase. The advantage given by the feedback process is clearly visible. 

\subsection{\label{sec22}Deterministic back-action and autonomous feedback}
In the previous section, we focused on measurement-based feedback, using an active classical controller that responds in real-time during the continuous measurement. We now look at setups where information extracted from the system is primarily discarded, however the induced back-action still drives the system to perform a desired operation.

We begin our discussion with autonomous control via Zeno dynamics, where the measurement strength approaches infinity. In this limit, measurement confines the system to specific subspaces, thereby affecting the dynamics within.
We then consider the more general case of autonomous feedback, sometimes termed coherent feedback. Here, a quantum system plays the role of a classical controller by driving the system to a target state or subspace. In both cases, the stochastic element of measurement is largely absent, even if one monitors the baths used to remove entropy. 

\begin{figure}
\begin{center}
{\includegraphics[width=1\linewidth]{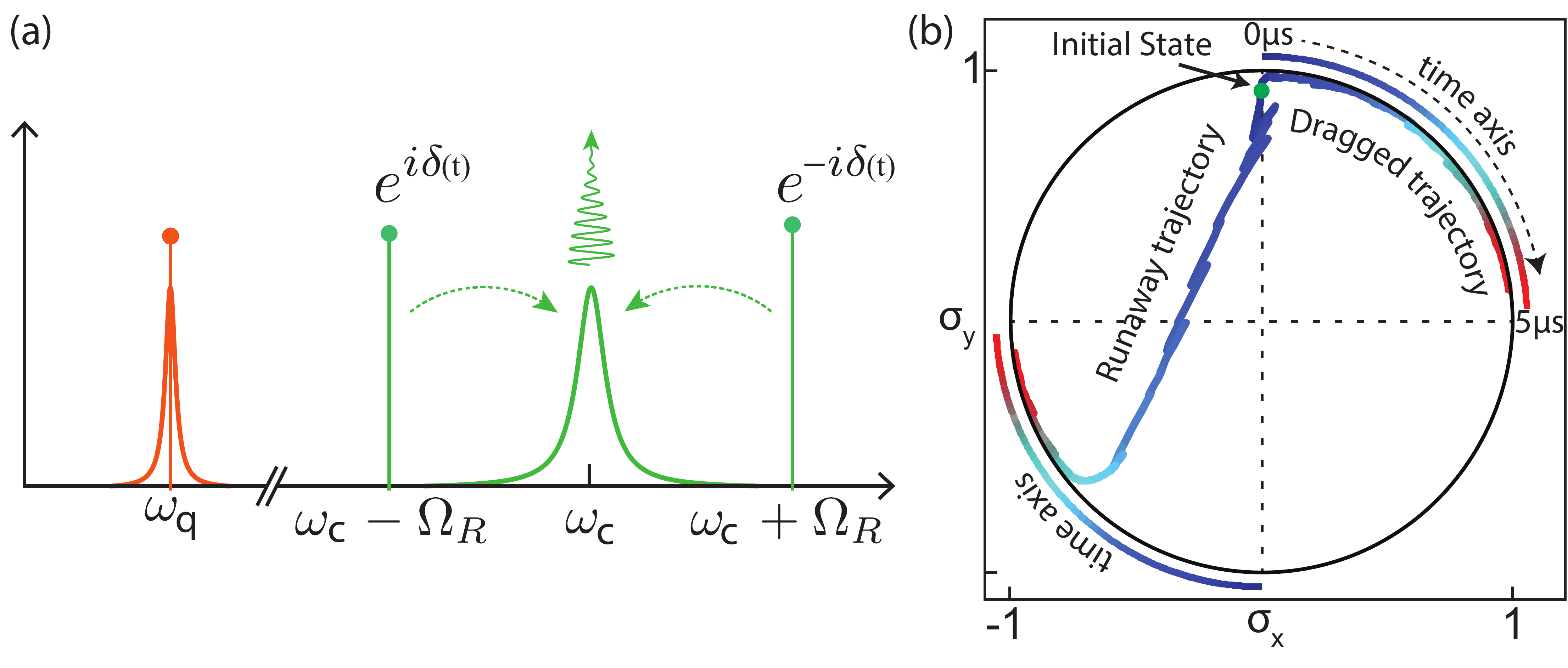}}
\caption{Dynamical control of measurement operator. (a) Autonomous feedback scheme used to create a variable measurement operator. The qubit is Rabi driven at a frequency $\Omega_R$ (orange). Additionally two cavity sidebands (green) detuned by $\Omega_R$ are applied. (b) Two example trajectories of a qubit, where the measurement operator was varied during the measurement, causing the trajectory to follow the operator pointer state. Trajectories are in the xy plane of the Bloch sphere of the effective qubit. The rotation rate is $v$=0.05MHz and a total duration of 5$\mu$s. Colors in the figure correspond to time evolution. The colored lines outside the Bloch sphere indicate the time axis going from blue for t=0$\mu$s to red for t=5$\mu$s, these illustrate the position of the measurement axis as function of time. The same colors correspond to the time evolution of the two trajectories shown. Figure adapted from \cite{hacohen-gourgy_incoherent_2017}}
\label{fig:ZenoDrag}
\end{center}
\end{figure}

\begin{figure}
\begin{center}
{\includegraphics[width=1\linewidth]{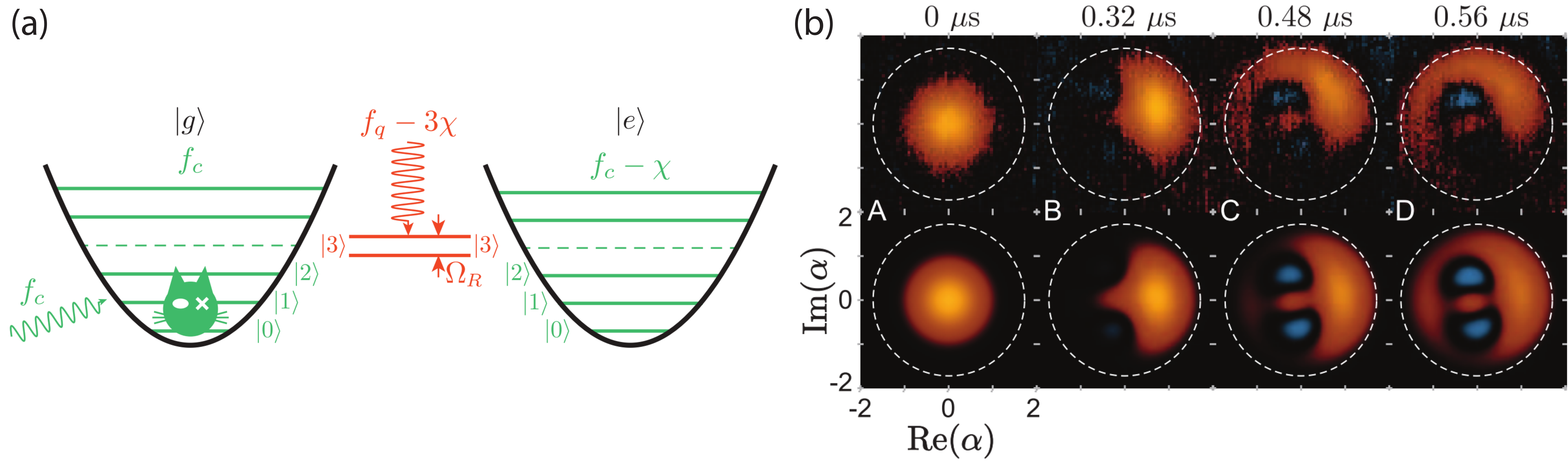}}
\caption{Zeno dynamics blockade of an oscillator. (a) Scheme illustration. Cavity energy levels dependent upon the qubit state (green). Blockade drive (orange). The blockade was created by driving the qubit transition corresponding to the 3rd photon number. The driving was at a rate $\Omega_R = 6.23$ MHz far larger than the linewidth $\gamma = 0.77$ MHz. In addition to the blockade, a coherent drive at the cavity frequency was applied. (b) Wigner function tomography as function of time. Figure adapted from Bretheau et al.\cite{bretheau_quantum_2015}}
\label{fig:ZenoDyn}
\end{center}
\end{figure}

\subsubsection{\label{sec222}Quantum Zeno dynamics}
The quantum Zeno effect describes the ability of measurements to freeze the dynamics of a system completely as the measurement strength goes to infinity. 
These dynamics arise since the back-action drives the system to a certain pointer state defined by the measurement operator\cite{misra_zenos_1977,pascazio_dynamical_1994}. A simple example with relevance to quantum control is the situation in which during the measurement, the measurement operator itself is varied over some time scale $1/\nu$. In the limit where $\Gamma_D / \nu \rightarrow \infty$, where $\Gamma_D$ is the measurement induced dephasing, the system will follow the pointer state deterministically. The rate of `escape' from the pointer state is $\nu^2 / \Gamma_D$. This situation was first suggested by Aharonov and Vardi\cite{aharonov_meaning_1980}, and was recently demonstrated experimentally in circuit-QED\cite{hacohen-gourgy_incoherent_2017}.

While conceptually simple, the experiment requires a technique for dynamically varying the measurement operator which is itself a form of autonomous feedback. We will discuss it in a more general context in the next subsection. To illustrate the dynamics, we show two example  trajectories selected from the experiment in Fig.\ref{fig:ZenoDrag}b. One trajectory illustrates successful dragging of the qubit state, whose state remains pure. The other, due to the finite measurement strength, undergoes a jump and continues to get dragged along the opposite pole. The described situation relates to the feedback scheme of Fig. \ref{fig:FBschemes}c. However, as can be seen, it is only truly deterministic in the limit $\Gamma_D / \nu \rightarrow \infty$. For finite measurement strength, it is possible to reduce the escape probability by adding feedback of the type shown in Fig. \ref{fig:FBschemes}b. In this situation, instead of varying the measurement operator deterministically, one uses the information from the detector to adjust the measurement operator. For a qubit, the operator needs to be continuously set to measure the axis exactly between the target state and the observer's best estimate of the current state of the qubit\cite{tanaka_robust_2012}. This protocol has been shown to be optimal\cite{martin_quantum_2020}. Note that although measurement-based feedback improves the success rate, the protocol achieves approximately deterministic control even if one does not monitor the measurement outcome.

In the previous example above the measurement acquired information about the whole system, a qubit in that case. However, this does not have to be the case, and the measurement can be applied to certain subspaces only. In such a situation, termed quantum Zeno dynamics\cite{misra_zenos_1977,facchi_quantum_2008}, the system dynamics are confined into the subspaces defined by the measurement. The dynamics remain unitary within each degenerate subspace of the measurement operator, but are strongly modified by measurement-induced confinement. This effect is gaining increased attention for applications of quantum control\cite{schafer_experimental_2014,raimond_quantum_2012}. 

An instructive example of Zeno dynamics is the work by Bretheau et al.\cite{bretheau_quantum_2015}. Here the setup is again of a transmon in a cavity. Crucially $\chi \gg \kappa$. In this so called number split regime\cite{schuster_resolving_2007}, the qubit transition frequency is split into separated transitions, each corresponding to a different photon number in the cavity. In the experiment, a Zeno blockade is created at a selected Nth level of the cavity mode. Under a coherent drive applied to the cavity, the cavity displaces in the expected fashion up until the point where the coherent state reaches the radius corresponding to the Nth level, as shown in Fig. \ref{fig:ZenoDyn}. At that time point the coherent state spreads and a cat-like state is formed. In contrast to a driven, linear cavity (which remains in a Gaussian state at all times), the Wigner function takes on negative values, indicating non-classical behavior. The emergence of non-classical dynamics is a manifestation of the fact that Zeno dynamics can transform almost any trivial quantum system into one in which universal control is possible, \textit{i.e.} any state state can be prepared\cite{burgarth_exponential_2014}. 

\subsubsection{\label{sec221}Autonomous feedback}

Sideband cooling is among the earliest forms of autonomous feedback, although is was not presented as a form of feedback at the time\cite{wineland_proposed_1975}. 
In sideband cooling, the controlled system coherently exchanges energy with a secondary quantum system (the controller) which is continually reset via dissipation. 
Let us now see how this is carried out in our transmon in a cavity system, as shown by Murch et al.\cite{murch_cavity-assisted_2012}. The protocol illustrates how the coherent feedback enables control over the dissipation rate and enables preparation of highly non-thermal states despite its reliance on dissipation. The protocol begins by Rabi driving a transmon qubit at $\Omega_\mathrm{R} \gg \kappa$, which creates an effective low-frequency qubit with dressed states $|\pm\rangle \equiv (|e\rangle \pm i|g\rangle)/\sqrt{2}$. Simultaneously, a cavity sideband is applied at $\omega_\mathrm{cav}-\Omega_\mathrm{R}$, where $\omega_\mathrm{cav.}$ is the cavity resonance frequency. The interaction is proportional to the Hamiltonian,
\begin{align}
\label{eq:HCoolingeff}
H_\text{eff} &\propto 
a \sigma^\dagger e^{i \delta} + a^\dagger \sigma e^{-i \delta}
\end{align}
where $\delta$ is a phase set by the sideband. The sideband, drives a transition from the $|-,0\rangle$ state to $|+,1\rangle$ state, where $0,1$ count the number of photons in the cavity. The cavity then decays, emitting a photon and leaving the system in the $|+,0\rangle$ state, which is not affected by the sideband. In our paradigm, the cavity can be understood as the quantum controller, and the sideband couples the qubit to the controller. The sideband amplitude sets the rate at which the feedback controller can respond, yet it is limited by the cavity linewidth $\kappa$ which is the maximum rate of entropy evacuation.

Photon emission during the cooling process clears the excess entropy. Notably, this photon carries information about the state of the qubit before the emission. If one were to put a single photon detector, a detection event tells the observer that the qubit was in the $|-\rangle$ state; whereas no-detection increases the probability that the qubit was in the $|+\rangle$ state. Information about the phase of the photon can be obtained using the adaptive phase measurement that we discussed in the previous section. Thus coherent feedback not only enables autonomous state preparation, but also provides a method to engineer non-trivial measurements of a quantum system.

An even more striking example of measurement engineering arises when one considers the applications of two sidebands, as illustrated in Fig. \ref{fig:ZenoDrag}a. Here the interference of the photons scattered by each sideband process determines the encoded information. In the case where the sidebands are of equal amplitude and detuned above and below the cavity frequency by $\Omega_R$.  Eq. (\ref{eq:DispHam}) is transformed into the effective Hamiltonian\cite{hacohen-gourgy_quantum_2016}
\begin{align}
\label{eq:HSQMeff}
H_\text{eff} &= \frac{\chi \bar{a}_0}{2} \Big[ \underbrace{a \sigma^\dagger e^{i \delta} + a^\dagger \sigma e^{-i \delta}}_\text{cooling} + \underbrace{a \sigma e^{-i \delta} + a^\dagger \sigma^\dagger e^{i \delta}}_\text{anti-cooling} \Big] \nonumber \\
&=\frac{\chi \bar{a}_0}{2} (a+a^\dagger) \sigma_{\delta}
\end{align}
where $\chi$ is the strength of the dispersive qubit-cavity coupling from Eq. \ref{eq:DispHam}, $\bar{a}_0$ is the amplitude of each sideband, and $\sigma_\delta \equiv \sigma_x \cos \delta +\sigma_y \sin\delta$. In this scheme the controller (the cavity) is coupled to the qubit through two channels (the sidebands). The controller's coherence plays a significant role, allowing interference  between the photons before they exit the cavity. The sideband's relative phase $\delta$ determines the interference between the up- and down-converted photons, which in turn dictates which qubit observable $\sigma_\delta$ is encoded in the signal. 

The resulting Hamiltonian is a resonant cavity drive, the sign of which depends on the qubit state along the $\sigma_\delta$ axis. Detecting the cavity output field with quantum efficiency $\eta$ yields a measurement of the qubit at a rate $\Gamma_D \eta = 2 \chi^2 \bar{a}_0^2 \eta/\kappa$ in the $\sigma_\delta$ basis, which can be dynamically controlled. This is the technique used in the Zeno example in the previous section to vary the measurement operator. 

The above technique can be generalized from a qubit to an N-level system. For an N-level system, there are up to N(N + 1)/2 distinct transitions, each of which can be driven and coupled to the cavity via sidebands. Although the generalization is more complicated than the qubit case, it can nevertheless be handled analytically. The end result is that far more measurement operators may be produced than in the qubit case. We refer the interested reader to\cite{martin_quantum_2020}.

The above autonomous feedback concepts have also been applied to multiqubit systems to create non-trivial states, such as stabilize selected states in a chain of 3 transmons in a cavity\cite{hacohen-gourgy_cooling_2015}, stabilize entanglement of two qubits in two coupled cavities through symmetry selecting rules\cite{kimchi-schwartz_stabilizing_2016}, and stabilizing entanglement of two qubits in the same cavity\cite{shankar_autonomously_2013}. 

\begin{figure}
\begin{center}
{\includegraphics[width=1\linewidth]{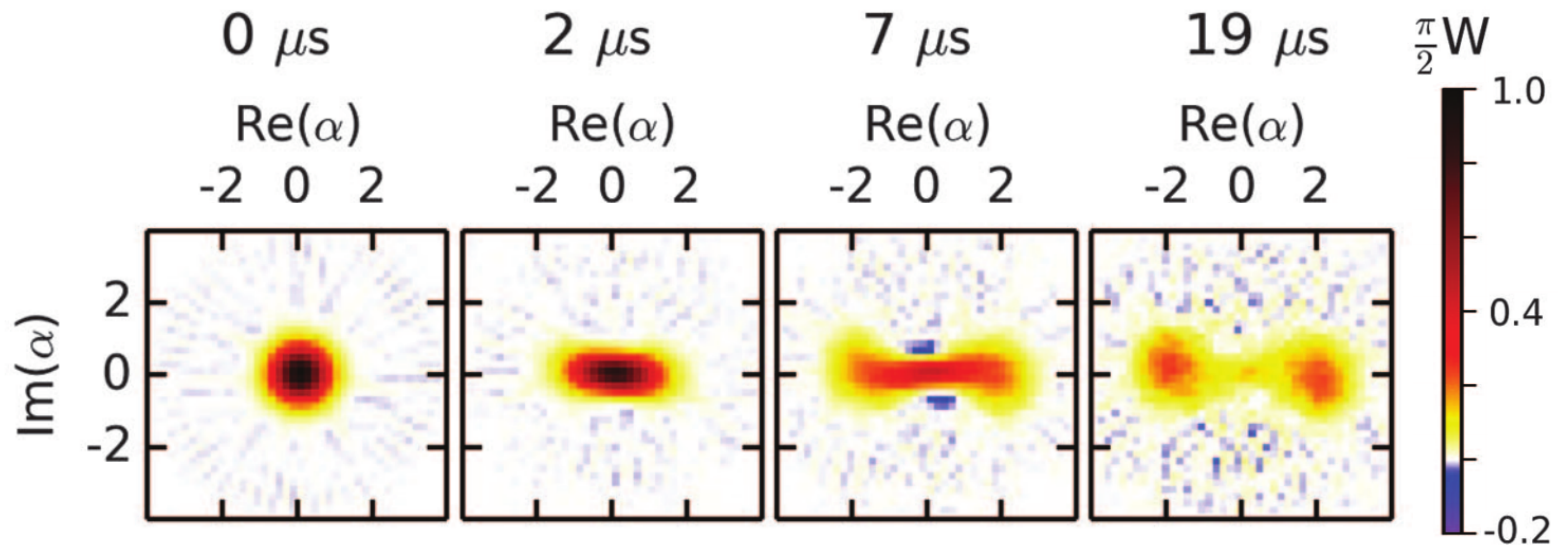}}
\caption{Evolution as function of time of the cavity state. Reconstructed Wigner function at 4 time points. Cavity starts in vacuum state ($t=0 \mu s$), then stabilizing drive is applied. First a squeezed like state forms due to the 2 photon pump ($t=2 \mu s$), then the cat state is formed after 7 $\mu s$. The $|\pm \alpha \rangle$ subspace is stabilized, however after a while ($t=19 \mu s$) the cat state has lost its coherence. This is evident by the lack of coherence fringes in the Wigner function. Figure adapted from Leghtas et al.\cite{leghtas_confining_2015}}
\label{fig:StabCat}
\end{center}
\end{figure}

Quantum error correction (QEC) is one of the most highly-sought after and challenging applications of quantum feedback. In QEC, one uses extra degrees of freedom to redundantly encode quantum information in a non-local fashion. In this way, deformations to the logical state resulting from local noise can be corrected without revealing the encoded information. In general, the larger the redundancy, the more coherence times can be extended. The extra resources for creating the redundancy required for a target protection level can be extremely large. The cavity in our system is a harmonic oscillator, which theoretically has an infinite Hilbert space. Thus, one path to overcome the resource overhead is to invert the role of the qubit and oscillator. The quantum information is then encoded in the large Hilbert space of the oscillator instead of the qubit, and the qubit is now the ancillary system that allows one to perform non-linear operations in the oscillator Hilbert space. Aside from the obvious advantage of a large Hilbert space, the use of harmonic oscillators also permits circumvention of several key no-go theorems in quantum error correction, which could drastically reduce resource overhead\cite{guillaud_repetition_2019}.

One of the more studied and promising encodings in this respect are the cat codes\cite{leghtas_hardware-efficient_2013,leghtas_deterministic_2013,mirrahimi_dynamically_2014,guillaud_repetition_2019}. Here, a logical qubit is encoded in a continuous-variable state called cat-state. Most relevant for our review is the fact that these cat-states can be continuously pumped and coupled to a reservoir to stabilize them\cite{puri_stabilized_2019,puri_engineering_2017,mirrahimi_dynamically_2014,guillaud_repetition_2019}. This corresponds to the feedback scheme of Fig. \ref{fig:FBschemes}c. In what follows, we introduce the cat states and the autonomous feedback stabilization of these logical qubits, and then elaborate on the latest advances of this approach.

Cat-states are defined as superpositions of coherent states $|\alpha\rangle$. A logical space of size two can be spanned using $|\alpha\rangle$, $|-\alpha\rangle$ and their superpositions. $|C \pm \rangle = |\alpha \rangle \pm |-\alpha\rangle$ are referred to as even (+) and odd (-) cat-states, corresponding to their number state parity. A logical qubit can be defined for example as $|0_L\rangle = |\alpha\rangle$ and $|1_L\rangle = |-\alpha\rangle$. Importantly, this subspace can be stabilized by continuously pumping and removing pairs of photons. This has been proposed theoretically both for a squeezing Hamiltonian with a two photon dissipator and a Kerr-nonlinear type Hamiltonian with a 2-photon pump, and demonstrated experimentally in circuit-QED\cite{lescanne_exponential_2019,leghtas_confining_2015,grimm_kerr-cat_2019}. Like many applications unique to circuit-QED, it is the strong non-linearity that enables physical realization. We introduce the Kerr-nonlinear Hamiltonian that stabilizes the cat-state subspace,
\begin{align}
\label{eq:Hkerr}
H_\text{Kerr} &= -K a^{\dagger 2} a^2 + \epsilon a^{\dagger 2} + \epsilon^{*} a^2
\end{align}
where $K$ is the Kerr nonlinearity and $\epsilon$ is the two-photon pump amplitude. This Hamiltonian has two degenerate ground states $|\pm \alpha \rangle$, with $\alpha=\sqrt{\epsilon / K}$. 

This Hamiltonian can be realized experimentally using a single transmon dispersively coupled to two cavity modes, a memory mode $\omega_m$ for storing the cat-state, and a fast mode $\omega_r$ for the engineered dissipation. The above interaction is achieved by adding a strong pump at a frequency $\omega_p=2 \omega_m- \omega_r$, and weak one at a frequency $\omega_r$. This pumping scheme generates the following effective Hamiltonian\cite{leghtas_confining_2015},
\begin{align}
\label{eq:HKerrCatExp}
H = g^{*} a_m^{2} a_r^\dagger + g a_m^{\dagger 2} a_r + \epsilon_d^{*} a_r + \epsilon_d a_r^\dagger \\ \nonumber - \chi_{ms} a_m^\dagger a_m a_r^\dagger a_r - \chi_{mm} a_m^{\dagger 2} a_m^{2} - \chi_{rr} a_r^{\dagger 2} a_r^{2}
\end{align}
where $ \epsilon_d $ is the weak resonant drive, and $g$ is proportional to the pump amplitude. The fast mode $r$ can be adiabatically eliminated, effectively resulting in Eq. \ref{eq:Hkerr} for the memory mode. Fig. \ref{fig:StabCat} shows data from Leghtas et al.\cite{leghtas_confining_2015}, where they applied this scheme. Higher order terms that cause an additional rotation in the cavity phase space have been dropped from Eq. \ref{eq:HKerrCatExp}, this effect can be seen in the data in Fig. \ref{fig:StabCat}, and can affect the logical encoding. The above method for obtaining $H_{Kerr}$ can be significantly improved by shunting the Josephson of the transmon with a large inductance\cite{lescanne_exponential_2019,grimm_kerr-cat_2019}, by placing a few larger Josephson junctions in parallel.

The feedback aspect arises from two attractors that arise at $|\pm\alpha\rangle$, and since the primary dissipation is a two-photon process, any local perturbation in the phase space of the oscillator is corrected without leaking logical information. The tunneling between the two ground states is exponentially suppressed with $|\alpha|^2$. Single photon loss under these dynamics does not take the state out of our subspace, however it does cause decoherence. Such events turn an even cat-state to an odd cat-state and vice versa, $a|C \pm \rangle = |C \mp \rangle$. This is seen in Fig. \ref{fig:StabCat}, where after a long enough time the cat-state has lost its coherence. This autonomous feedback stabilizes the cat-state subspace, and decreases the bit-flip rate exponentially, yet at the expense of increasing the phase-flip rate linearly with the cat size\cite{lescanne_exponential_2019}. This type of error correction is known as noise-biased. Recently a scheme for a fully fault-tolerant quantum computation based on these noise biased pumped cats has been proposed\cite{guillaud_repetition_2019}. It was shown that by concatenating the above scheme with a 1D repetition code, both errors can be corrected, and a universal set of protected gates can be constructed. This opens up a new path towards fault-tolerant quantum computation based on continuous measurements in superconducting circuits.

\section{\label{sec3}Summary and outlook}
We introduced here the basics of continuous measurements and feedback control, showing how they play out in the context of circuit-QED. This microwave quantum optics platform offers greatly increased flexibility compared to natural atomic systems, and offers interactions strength exceeding their natural counterparts by several orders of magnitude.
Furthermore, parametric amplifiers can be designed with high quantum efficiencies. These are the key abilities that make this platform ideal for continuous measurements and feedback control. One example of this versatility is the adaptive phase experiment we discussed, where several abilities were required in the same setup. To generate the photon with a specific mode shape, sideband cooling was employed, which is a simple form of autonomous feedback. In addition, a measurement based feedback scheme was used to optimize the detector measurement basis in real-time. The combined setup allowed to emit a photon in a controllable way, detect it's phase through feedback, and verify the system dynamics using the qubit. The thorough understanding of the role of measurements in a broad sense leads to more control options for quantum systems. Due to the versatility, circuit-QED has proved to be an ideal platform to explore such enhanced quantum control.

Quantum trajectories offer both an experimental tool and a unique theoretical perspective. Experiments up to now have shown that the master equation formalism introduced here for continuous measurement works well for extracting the quantum trajectories in Markovian systems. Experiments on multiple qubits and non-Markvian dynamics are a natural path of progression waiting to be explored. In terms of fluorescence, an interesting potential avenue would be the unravelling of more intricate AMO phenomena, such as the Mollow triplet\cite{toyli_resonance_2016}, or superradiance and superradiant lasing\cite{martin_quantum_2020}. More exotic directions would be exploring many body systems using continuous measurements\cite{yunger_halpern_quasiprobability_2018}; monitoring of single Fermion spins trapped in a Josephson junction\cite{hays_continuous_2019} and electrons trapped on superfluid helium\cite{yang_coupling_2016}.

We have seen that the enhanced capabilities afforded by the addition of feedback to continuous monitoring. While this comes across as primarily application-oriented, it also suggests interesting directions at the foundations level. One example is the recent observation of quantum jumps as a continuous process\cite{minev_catch_2019}, where feedback was used to stop the system midway through the quantum jump process. Such demonstrations enforce our already existing understanding that all measurement processes are continuous, and it is only a question of weather our detector can resolve the relevant timescales to unravel the dynamics.
Understanding back-action as a control resource leads to further interesting possibilities like feedback-induced quantum phase transitions due to the continuous monitoring process itself\cite{ivanov_feedback-induced_2020}. Also, continuous measurement and feedback open up certain simulation abilities, as has recently been shown in an atomic system\cite{munoz-arias_simulating_2020}.

Arguably, the ultimate and possibly most difficult goal of quantum feedback is to stabilize an encoded logical quantum state. Continuous measurements are a powerful tool for implementing these ideas\cite{atalaya_error_2019, ahn_continuous_2002}. One direction we touched upon in this review is the autonomous error correction schemes. These are true tour-de-force demonstrations of the circuit-QED system universal control, both in terms of coherent control, back-action engineering, and dissipative coupling to the environment. Specifically, the ability to engineer a two-photon pump and dissipation mechanism to stabilize a cat-state stands out in the example we discussed above. Future directions could look at scaling these concepts\cite{guillaud_repetition_2019}, or stabilizing more complex states such as GKP states.

\bibliography{references}

\end{document}